\documentclass[11pt]{article}
\usepackage{jheppub}

\newcommand\fft[2]{{\frac{#1}{#2}}}

\newcommand\nn{\nonumber}

\begin{document}

\preprint{MCTP-15-06 \begin{flushright} \vspace{-1em}
ITP-UU-15/03\end{flushright}}

\title{\boldmath High-Temperature Expansion of Supersymmetric Partition Functions}

\author[a]{Arash Arabi Ardehali,}
\author[a]{James T. Liu,}
\author[b]{and Phillip Szepietowski}

\affiliation[a]{Michigan Center for Theoretical Physics, Randall Laboratory of Physics,\\
The University of Michigan, Ann Arbor, MI 48109--1040, USA}
\affiliation[b]{Institute for Theoretical Physics \& Spinoza Institute, \\
Utrecht University, 3508 TD Utrecht, The Netherlands}

\emailAdd{ardehali@umich.edu} \emailAdd{jimliu@umich.edu}
\emailAdd{P.G.Szepietowski@uu.nl}

\abstract{Di~Pietro and Komargodski have recently demonstrated a
four-dimensional counterpart of Cardy's formula, which gives the
leading high-temperature ($\beta\rightarrow 0$) behavior of
supersymmetric partition functions $Z^{SUSY}(\beta)$. Focusing on
superconformal theories, we elaborate on the subleading
contributions to their formula when applied to free chiral and U(1)
vector multiplets.  In particular, we see that the high-temperature
expansion of $\ln Z^{SUSY}(\beta)$ terminates at order $\beta^0$. We
also demonstrate how their formula must be modified when applied to
SU($N$) toric quiver gauge theories in the planar
($N\rightarrow\infty$) limit. Our method for regularizing the
one-loop determinants of chiral and vector multiplets helps to
clarify the relation between the 4d $\mathcal{N}=1$ superconformal
index and its corresponding supersymmetric partition function
obtained by path-integration.}

\maketitle \flushbottom

\section{Introduction}

Some time ago, Cardy famously employed modular invariance to obtain
the high-temperature behavior of conformal field theory (CFT)
partition functions in two dimensions \cite{Cardy:1986ie}. This
result has since been exploited in a variety of contexts, including
the statistical physics of black holes
\cite{Strominger:1996,Strominger:1998,Hartman:2014}. Cardy's formula
gives the leading order divergence of the CFT partition function
$Z(\beta)$ as the inverse temperature $\beta$ goes to zero:
\begin{equation}
\ln Z(\beta)\sim \frac{\pi^2 c_L}{6\beta}.\label{eq:Cardy}
\end{equation}
Here, $c_L$ is the left-handed CFT central charge, and we have
focused on the holomorphic sector for simplicity. Note that the term
``Cardy formula'' is often applied to the expression, derived from
the above relation, for the micro-canonical entropy of a 2d CFT at
high energies. However, in the present work, by ``Cardy formula'' we
always refer to the above canonical version for the asymptotic
high-temperature expansion of $\ln Z$.

Similar formulae had been long sought in higher dimensions without
much success, partly because Cardy's main tool, modular invariance,
has no known higher-dimensional counterpart. Recently, Di~Pietro and
Komargodski have combined ideas from supersymmetry and hydrodynamics
\cite{Banerjee:2012iz,Jensen:2012jh} to obtain the high-temperature
behavior of supersymmetric (SUSY) partition functions in four and
six dimensions \cite{DiPietro:2014}. Here we expand on their result
in the context of four-dimensional superconformal field theories
(SCFTs).

By SUSY partition function we mean the one computed with periodic
boundary conditions for fermions along the thermal circle; this
amounts to an insertion of $(-1)^{F}$ when the partition function is
represented as a weighted sum over the states, and makes it
independent of exactly marginal couplings \cite{Witten:1982}.
Therefore, one might anticipate that the partition function displays
universal high-temperature behavior depending only on the 4d central
charges.  This was realized by Di~Pietro and Komargodski, who
demonstrated the relation \cite{DiPietro:2014}
\begin{equation}
\ln Z^{SUSY}(\beta)\approx \frac{16\pi^2
(c-a)}{3\beta},\label{eq:DK}
\end{equation}
where $c$ and $a$ are the central charges of the 4d SCFT, and where
the spatial manifold is taken to be the round $S^3$. (In the main
text we focus on the round $S^3$, while relegating the case of the
squashed sphere to appendix \ref{app:A}.)

The formula (\ref{eq:DK}) can be thought of as the leading order
result in a high-temperature expansion. In this paper we explore the
subleading corrections to Eq.~(\ref{eq:DK}) and provide evidence
that it receives only ``non-perturbative'' corrections in $\beta$
(of the type $e^{-1/\beta}$), and $\mathcal{O}(\log\beta)$ and
$\mathcal{O}(\beta^0)$ corrections. While these corrections have
already been pointed out by Di~Pietro and Komargodski in
\cite{DiPietro:2014}, we conjecture that the series expansion of
$\ln Z^{SUSY}(\beta)$ around $\beta=0$ terminates at $\mathcal
O(1)$, and that no corrections arise at order $\beta$ or higher.

To explore the subleading behavior of SUSY partition functions, it
proves helpful to understand the relation between their
path-integral representation and their representation as a weighted
sum. The latter is called the superconformal index
\cite{Romelsberger:2005eg,Kinney:2005ej}, and may be defined with
two fugacities as
\begin{equation}
\mathcal
I(p,q)=\mathrm{Tr}\left[(-1)^Fe^{-\hat\beta(\Delta-2j_2-\fft32r)}
p^{j_1+j_2+\fft12r}q^{-j_1+j_2+\fft12r}\right].
\label{eq:IndexTwofugDef}
\end{equation}
Here the trace is over the Hilbert space of the SCFT in radial
quantization, $\Delta$ is the conformal dimension of the state, $r$
is its R-charge, and $(j_1,j_2)$ are its
$\mathrm{SO}(4)=\mathrm{SU}(2)_1\times\mathrm{SU}(2)_2$ quantum
numbers. Only states with $\Delta-2j_2-\fft32r=0$  contribute to the
index, so it is independent of $\hat{\beta}$. The index may be
related to the partition function on the round $S^3\times S^1$ by
taking $p=q=e^{-\beta},$ where $\beta$ is identified with the radius
of the $S^1.$  We thus have
\begin{equation}
\mathcal I(\beta)=\mathcal
I(e^{-\beta},e^{-\beta})=\mathrm{Tr}\left[(-1)^F
e^{-\beta(\Delta-\fft12r)}\right]. \label{eq:IndexDef}
\end{equation}
The generalization to squashed 3-sphere (and therefore non-equal
fugacities in the index) will be discussed in appendix \ref{app:A}.

In the following, we will refer to $Z^{SUSY}$ obtained by
path-integration as the ``SUSY partition function'', and to
$\mathcal{I}$ as the ``index''. The relation between these two
quantities is
\cite{Kim:2012ava,Closset:2014,Assel:2014}%
\footnote{See Eq.~(\ref{eq:ZandIgenSqca}) for the relation between
the index with two fugacities and the SUSY partition function on the
squashed 3-sphere as spatial manifold.}
\begin{equation}
\mathcal{I}(\beta)=e^{\beta
E_{\mathrm{susy}}}Z^{SUSY}(\beta)=e^{\frac{4(3c+a)}{27}\beta}Z^{SUSY}(\beta),
\label{eq:ZandIgen}
\end{equation}
where the supersymmetric Casimir energy
\cite{Kim:2012ava,Assel:2014,Lorenzen:2014} is given by
$E_{\mathrm{susy}}=4(3c+a)/27$.  It is worth noting that in
\cite{Closset:2014,Assel:2014}, an extra $\mathcal{O}(1/\beta)$
factor was present in the exponent of the prefactor. However, as
mentioned in \cite{DiPietro:2014,Lorenzen:2014}, and as highlighted
below, an alternative regularization of the computations in
\cite{Closset:2014,Assel:2014,Lorenzen:2014} would eliminate that
extra factor.

The relation (\ref{eq:ZandIgen}), when combined with our claim that
$\ln Z^{SUSY}(\beta)$ has no $\mathcal{O}(\beta)$ term in its
asymptotic high-temperature expansion, implies an
$\mathcal{O}(\beta)$ term (namely $\frac{4(3c+a)}{27}\beta$) in the
high-temperature expansion of $\ln\mathcal{I}(\beta)$. This was
conjectured in \cite{Ardehali:2015}.

The fact that the path-integral and the trace representation of the
partition function are related by anomaly-dependent factors has a
well-known counterpart in 2d CFT which we will review below.

Also well-known for 2d CFTs is the breakdown (except for sparse CFTs
\cite{Hartman:2014}) of Cardy's formula in the limit of large
central charge. The analogous situation for the
Di~Pietro-Komargodski formula (\ref{eq:DK}) was noted in
\cite{Ardehali:2014} from a case-by-case study of some holographic
SCFTs, and also the $A_k$ SQCD fixed points. It was observed that in
the planar limit, which is the 4d gauge theory counterpart of the 2d
large-$c$ limit, the 4d index has a rather non-universal
high-temperature behavior which is not dictated solely by the
central charges. In this paper we systematically investigate
large-$N$ toric quivers, and see the modification of (\ref{eq:DK})
explicitly. This modification can be intuitively understood as a
kind of non-commutativity between the high-temperature limit and the
planar limit.

In appendix \ref{app:A}, we generalize our computations to the case
with squashed 3-sphere as the spatial manifold.  Among other things,
we derive a powerful identity (given in Eq.~(\ref{eq:GammaFVSq}),
generalizing Eq.~(\ref{eq:GammaFV})) which relates the elliptic
Gamma function to the non-compact quantum dilogarithm, and makes the
high-temperature behavior of the index of a chiral multiplet quite
transparent and its connection with the 3d partition function
manifest.

The results of appendix \ref{app:A} will be employed in appendix
\ref{app:B} to demonstrate the relation between the high-temperature
expansion of the index and the holographically derived prescriptions
of \cite{Ardehali:2015} for extracting the central charges from the
single-trace index. In particular, we show that the prescriptions of
\cite{Ardehali:2015} probe only the $\mathcal{O}(\beta)$ term in the
high-temperature expansion of $\ln\mathcal{I}$, and are insensitive
to the leading $\mathcal{O}(1/\beta)$ behavior.

The organization of this paper is as follows. In the next section,
for the purpose of orientation and also to highlight later some
analogies with 4d SCFTs, we review Cardy's formula and the
subleading corrections it receives in a high-temperature expansion.
In section \ref{sec:3} we consider free chiral and U(1) vector
multiplets and examine the subleading corrections to the
Di~Pietro-Komargodski formula. Section \ref{sec:4} contains the
discussion of large-$N$ toric quivers, and our concluding remarks
are presented in the last section. SCFTs with squashed three-sphere
as their spatial manifold are treated in appendix \ref{app:A}, and
in appendix \ref{app:B} the connection between our findings in the
present paper and the proposals of \cite{Ardehali:2015} are
clarified.

\section{Cardy's formula for 2d CFTs}\label{sec:2}

Before proceeding to four dimensions, we review some well-known
facts about Cardy's formula for 2d CFTs. Cardy's formula
\cite{Cardy:1986ie} is obtained using the modular invariance of 2d
CFT partition functions
\begin{equation}
Z_{PI}(\tau)=Z_{PI}(-1/\tau),\label{eq:modular2d}
\end{equation}
where $-2\pi i\tau=\beta$, and we have added a subscript $PI$ since
we assume the partition function is computed by a path integral. For
simplicity we focus on the holomorphic sector of the CFT. If the
theory has a gapped spectrum, with its lightest state having energy
$\Delta_0$ with respect to the vacuum, the low-temperature
($\beta\rightarrow\infty$) partition function is dominated by the
vacuum contribution $e^{c_L\beta/24}$, while the next contribution
is down by a factor $e^{-\beta \Delta_0}$. Using
(\ref{eq:modular2d}) we arrive at Cardy's relation in
(\ref{eq:Cardy}), which we rewrite as
\begin{equation}
\ln Z_{PI}(\beta)\sim \frac{\pi^2 c_L}{6\beta}.\label{eq:simCardy}
\end{equation}
This formula leads to the micro-canonical entropy
$2\pi\sqrt{c_L(L_0-c_L/24)/6}$, which matches that of the
Strominger-Vafa black hole \cite{Strominger:1996}.

The subleading correction to (\ref{eq:simCardy}) is down by a factor
$e^{-4\pi^2 \Delta_0/\beta}$, which is non-analytic in $\beta$. In
other words Cardy's formula is correct to all orders in a
high-temperature expansion, and only receives ``non-perturbative''
corrections in $\beta$. Below, we will repeatedly use the symbol
$\sim$ to denote such all-orders equalities.

To highlight the analogy with 4d SCFTs we now assume the 2d CFT has
a single left-handed conserved U(1) current $J$ whose Laurent modes
satisfy the following commutation relations
\begin{equation}
\begin{split}
[L_n,J_m]&=-mJ_{m+n},\\
[J_m,J_n]&=2k m\delta_{m+n,0}, \label{eq:JcommRel}
\end{split}
\end{equation}
where $L_n$ are the Laurent modes of the energy-momentum tensor. For
example, a $(2,0)$ SCFT has such a conserved current with $k=c_L/6$,
sometimes referred to as the R-current.

Adding a chemical potential $\mu=2\pi iz/\beta$ for the U(1) charge
of the CFT states, one can define the following grand-canonical
partition function
\begin{equation}
\mathcal{I}_{2d}(\tau,z)=
\mathrm{Tr}(q^{L_0}y^{J_0}),\label{eq:GCpf}
\end{equation}
where $q=e^{2\pi i \tau}$, $y=e^{2\pi iz}$, and we have considered a
holomorphic CFT (with $c_R=0$) for simplicity. This partition
function is not modular invariant, but its path-integral
representation $Z_{PI}(\tau,z)$ is invariant under the modular
transformation $\{\tau,z\}\rightarrow\{-1/\tau,z/\tau\}$. The two
are related via \cite{Kraus:2007}
\begin{equation}
\mathcal{I}_{2d}(\tau,z)= e^{-i\pi k\frac{z^2}{\tau}+2\pi
i\tau\frac{c_L}{24}}Z_{PI}(\tau,z).\label{eq:2dIvsZ}
\end{equation}
Hence, modular invariance of $Z_{PI}$ can be combined with the
assumption that $\mathcal{I}_{2d}$ is dominated at low temperatures
by the vacuum, to give the high-temperature behavior
\begin{equation}
\mathcal{I}_{2d}(\tau,z)\rightarrow e^{-4\pi^2 k
\frac{z^2}{\beta}-\frac{c_L}{24}\beta}e^{\frac{\pi^2
c_L}{6\beta}}=e^{\mu^2 k \beta-\frac{c_L}{24}\beta}e^{\frac{\pi^2
c_L}{6\beta}}.\label{eq:2dCardyI}
\end{equation}
For $(2,0)$ SCFTs with $k=c_L/6,$ this variant of Cardy's formula
yields the micro-canonical entropy
$2\pi\sqrt{c_L(L_0-c_L/24)/6-J_0^2/4}$, which reproduces%
\footnote{Note that the actual derivation of this result is a bit
more subtle \cite{Breckenridge:1997}, as it relies not on $c_R = 0$,
but on taking the right-handed sector to be in a Ramond ground
state.}
the Bekenstein-Hawking entropy of spinning generalizations of the
Strominger-Vafa black hole \cite{Breckenridge:1997,Kraus:2007}.

For future reference, note that the $\mathcal{O}(\beta)$ term in the
high-temperature expansion
\begin{equation}
\ln \mathcal{I}_{2d}(\tau,z)\sim \frac{\pi^2 c_L}{6\beta} + \mu^2 k
\beta-\frac{c_L}{24}\beta.\label{eq:2dCardyImu1}
\end{equation}
differs from the one in
\begin{equation}
\ln Z_{PI}(\tau,z)\sim \frac{\pi^2 c_L}{6\beta}+ \mu^2\frac{k}{2}
\beta.\label{eq:2dCardyZmu1}
\end{equation}
In particular, when $\mu=0$, the difference is entirely due to the
familiar $-c_L/24$ Casimir energy on the torus; this Casimir energy is
computed by $Z_{PI}$, but not taken into account in our definition
of $\mathcal{I}_{2d}$ in (\ref{eq:GCpf}).

Finally, we remind the reader that Cardy's formula
(\ref{eq:simCardy}) fails for general 2d CFTs in which the
$c\rightarrow \infty$ limit is taken before (or at the same time as)
the $\beta\rightarrow 0$ limit; the asymptotic high-temperature
expansion may clash with (\ref{eq:simCardy}). This is because a
given large-$c$ CFT may have too many light states so that its
low-temperature partition function is no longer dominated by the
vacuum contribution alone. CFTs with a ``sparse'' spectrum of low
lying states avoid this breakdown. For related recent discussions
see \cite{Hartman:2014,Belin:2014,Hael:2014}. Similar sparseness
conditions for the grand-canonical partition function
(\ref{eq:GCpf}) remain to be formulated.

\section{Subleading corrections to the Di~Pietro-Komargodski
formula}\label{sec:3}

We now return to four-dimensions and explore the subleading
corrections to the Di~Pietro-Komargodski result, (\ref{eq:DK}).  We
claim that the high-temperature expansion of the SUSY partition
function on the round $S^3\times S^1$ has the form
\begin{equation} \ln Z^{SUSY}(\beta)\sim
\frac{16\pi^2(c-a)}{3\beta}-4(2a-c)\ln(\beta/2\pi)+\ln
Z_{3d},\label{eq:conjZ}
\end{equation}
which terminates at $\mathcal O(\beta^0)$, and is exact up to
non-analytic terms of the type $e^{-1/\beta}$. Here $Z_{3d}$ is the
supersymmetric partition function of the dimensionally reduced
theory on $S^3$, which in favorable cases can be computed by
localization \cite{Kapustin:2010,Jafferis:2012,Hama:2011}. Using
(\ref{eq:ZandIgen}), the above relation leads to the expansion of
the index
\begin{equation}
\ln \mathcal{I}(\beta)\sim
\frac{16\pi^2(c-a)}{3\beta}-4(2a-c)\ln(\beta/2\pi)+\ln
Z_{3d}+\frac{4(3c+a)\beta}{27}.\label{eq:conjI}
\end{equation}
The linear term on the RHS of (\ref{eq:conjI}) was conjectured in
\cite{Ardehali:2015}, based on holographically derived relations
between central charges and the index. That the high-temperature
expansions of $\ln\mathcal{I}(\beta)$ and $\ln Z^{SUSY}(\beta)$
differ only by an $\mathcal{O}(\beta)$ term is somewhat analogous to
the 2d story sketched in Eqs.~(\ref{eq:2dCardyImu1}) and
(\ref{eq:2dCardyZmu1}).

Upon squashing the $S^3$, the above relations are generalized to
(\ref{eq:logIsquashedCA}) and (\ref{eq:logZsquashedCA}). An
important feature arises in the expansion of the index in
(\ref{eq:logIsquashedCA}), that we would like to highlight already:
the linear term in $\beta$ encodes \emph{two linear combinations} of
$a$ and $c$, separated by their different dependence on the
squashing parameter. Therefore both central charges can be distilled
at high temperatures from the $\mathcal{O}(\beta)$ term of
$\ln\mathcal{I}$. In appendix \ref{app:B} we will demonstrate that
this observation is essential for making contact with the proposals
of \cite{Ardehali:2015}.

We now provide support for the claims (\ref{eq:conjZ}) and
(\ref{eq:conjI}) by investigating free chiral and U(1) vector
multiplets.  Of course, a general proof would require studying
non-abelian gauge theories and going beyond the free cases.

\subsection{Free chiral multiplet}\label{subsec:freeChi}

Consider now the concrete case of a free chiral multiplet of
off-shell R-charge $R$ (we say off-shell because a free chiral
multiplet is known to have R-charge 2/3 ``on-shell''). Its index can
be written as \cite{Dolan:2008}
\begin{equation}
\mathcal I_\chi(R,p,q)=\Gamma(z,\tau,\sigma), \label{eq:Ichi}
\end{equation}
where $p=e^{2\pi i\tau}$, $q=e^{2\pi i\sigma}$, and
$z=(\tau+\sigma)R/2$.  Here $\Gamma(z,\tau,\sigma)$ is the elliptic
Gamma function defined by
\begin{equation}
\Gamma(z,\tau,\sigma)=\prod_{j,k\ge0}\frac{1-e^{2\pi
i((j+1)\tau+(k+1)\sigma-z)}}{1-e^{2\pi
i(j\tau+k\sigma+z)}}.\label{eq:ellGamma}
\end{equation}
In order to investigate the high-temperature limit of
(\ref{eq:Ichi}), we make use of the SL(3,$\mathbb{Z}$) modular
property \cite{Felder:1999}
\begin{equation}
\Gamma(z,\tau,\sigma)=e^{-i\pi
M(z;\tau,\sigma)}\frac{\Gamma(\frac{z}{\tau},-\frac{1}{\tau},\frac{\sigma}{\tau})}{\Gamma(\frac{z-\tau}{\sigma},-\frac{1}{\sigma},-\frac{\tau}{\sigma})},
\label{eq:modprop}
\end{equation}
where
\begin{equation}
M(z;\tau,\sigma)=\fft{z^3}{3\tau\sigma}-\fft{\tau+\sigma-1}{2\tau\sigma}z^2
+\fft{\tau^2+\sigma^2+3\tau\sigma-3\tau-3\sigma+1}{6\tau\sigma}z+\fft1{12}(\tau+\sigma-1)(\tau^{-1}+\sigma^{-1}-1).
\end{equation}
This is analogous to how the SL(2,$\mathbb{Z}$) properties of 2d
partition functions allow a Cardy-type analysis, as briefly sketched
in the previous section.

Restricting to the case of equal fugacities ({\it i.e.}\ the round
$S^3$), the index can be written as
\begin{equation}
\mathcal{I}_\chi(R,\beta)=\Gamma(R\tau,\tau,\tau),\label{eq:IwithGamma}
\end{equation}
where $\tau=i\beta/2\pi$.  In order to study its $\tau\to0$ limit,
we resort to Theorem 5.2 of \cite{Felder:1999} which is derived from
(\ref{eq:modprop}), along with some straightforward manipulation, to
rewrite $\Gamma(R\tau,\tau,\tau)$ as
\begin{equation}
\Gamma(R\tau,\tau,\tau)=\frac{e^{-i\pi
M(\tau,R)}}{\psi(-(R-1))}\prod_{n=1}^{\infty}\frac{\psi\left(\frac{n+(R-1)\tau}{\tau}\right)}
{\psi\left(\frac{n-(R-1)\tau}{\tau}\right)},\label{eq:GammaFV}
\end{equation}
where
\begin{equation}
M(\tau,R)=\left(\frac{R-1}{6}\right)\fft1\tau+\left(\frac{R^2}{2}-R+\frac{5}{12}\right)
+\left(\frac{R^3}{3}-R^2+\frac{5R}{6}-\frac{1}{6}\right)\tau.\label{eq:M}
\end{equation}
The $\psi$ functions present on the RHS of Eq.~(\ref{eq:GammaFV})
can be expressed as
\begin{equation}
\ln\psi(R)=R\ln(1-e^{-2\pi i R})-\frac{1}{2\pi i}Li_2(e^{-2\pi i
R}).\label{eq:psidef}
\end{equation}
$\psi(R)$ has a zero of order $j$ at $R=j$, and a pole of the same
order at $R=-j$, for $j\in\mathbb{Z}^{>0}$.

The reader familiar with the 3d localization literature may notice
that $\psi(R)$ is related to the function $\ell(R)$ that Jafferis
uses in \cite{Jafferis:2012} via
\begin{equation}
\ell(R)=\ln\psi(-R)+\fft{i\pi
R^2}2-\fft{i\pi}{12}.\label{eq:ell-Psi}
\end{equation}
$\ell(R)$ has the useful property that $\ell(-R)=-\ell(R)$. From the
information on poles and zeros of $\psi(R)$, we see that $\ell(R)$
is singular at $R\in\mathbb{Z}-\{0\}$. For future reference, we add
that $\ell(R)$ is related to the function $s_{b=1}(R)$ in
\cite{Hama:2011} (see also appendix \ref{app:A}) via
\begin{equation}
\ell(R)=\ln s_{b=1}(iR).\label{eq:ell-s}
\end{equation}

To obtain the high-temperature behavior of (\ref{eq:GammaFV}) we
utilize the fact that the $\psi$ function is exponentially close to
one when its argument has a large negative imaginary part
\cite{Felder:1999}. This means that in the limit $\beta=-2\pi
i\tau\rightarrow 0$, the infinite product in Eq.~(\ref{eq:GammaFV})
can be replaced with one, yielding
\begin{equation}
\ln\mathcal I_\chi(R,\beta)=\ln\Gamma(R\tau,\tau,\tau)\sim -i\pi
M(\tau,R)-\ln{\psi(-(R-1))}.\label{eq:GammaExp}
\end{equation}
Recall that $\sim$ means to all orders in a high-temperature
expansion; non-analytic corrections of the type $e^{-1/\beta}$
coming from the infinite product on the RHS of (\ref{eq:GammaFV})
are present but are not part of the perturbative expansion.
Substituting in (\ref{eq:M}) for $M(\tau,R)$ and making use of
(\ref{eq:ell-Psi}) then gives
\begin{equation}
\ln\mathcal{I}_\chi(R,\beta)\sim\frac{-\pi^2
(R-1)}{3\beta}+\ell(-(R-1))+\beta\left(\frac{R^3}{6}-\frac{R^2}{2}+\frac{5R}{12}-\frac{1}{12}\right),
\label{eq:IchiExp}
\end{equation}
in perfect agreement with the conjectured form of the index,
(\ref{eq:conjI}). Importantly, there are no terms of order $\beta^2$
or higher on the RHS. Also, since for a chiral multiplet $2a-c=0$,
there is no $\mathcal{O}(\log\beta)$ term here, unlike in the case
of a free U(1) vector multiplet \cite{DiPietro:2014}.

The temperature-independent term $\ell(-(R-1))$ in
(\ref{eq:IchiExp}) is precisely the log of the partition function of
a 3d chiral multiplet \cite{Jafferis:2012}; this is the well-known
result that the $\mathcal{N}=1$ 4d index reduces, as
$\beta\rightarrow 0$, to the 3d partition function, after (and only
after) its $\mathcal{O}(1/\beta)$ divergent exponent is removed.
Related discussions can be found in
\cite{Dolan:2011sv,Gadde:2012y,Imamura:2011,Niarchos:2012,Agarwal:2012,Aharony:2013}.
The argument above is, however, in our opinion the most transparent
derivation of the reduction result for a chiral multiplet (see also
appendix \ref{app:1} for the case with squashing).

Having established the expansion of the index, we now turn to the
high-temperature behavior of $Z^{SUSY}_\chi(R,\beta)$, the SUSY
partition function of a chiral multiplet with R-charge $R$. The
computation is done by KK compactification of the theory on the
thermal circle, calculating the contribution to the free energy of
the $n$-th KK modes $\ln Z_\chi^{(n)}(R,\beta)$, and then summing up
over $n$. A similar calculation was performed in the appendix of
\cite{DiPietro:2014} to obtain the leading high-temperature behavior
of the SUSY partition function. For $\ln Z_\chi^{(n)}(R,\beta)$, we
may use the results of \cite{Jafferis:2012,Hama:2011}
\begin{equation}
\ln Z_\chi^{(n)}(R,\beta)=\ln s_{b=1}(i-iR-\frac{2\pi
n}{\beta})=\ell(1-R+\frac{2\pi i n}{\beta}).\label{eq:nthKKchi}
\end{equation}
Summing over the KK tower now gives
\begin{equation}
\begin{split}
\ln
Z^{SUSY}_{\chi}(R,\beta)&=\sum_{n\in\mathbb{Z}}\ell(1-R+\frac{2\pi i
n}{\beta})\\
&=\ell(-(R-1))+\sum_{n>0}[\ell(1-R+2\pi i n/\beta)-\ell(-1+R+2\pi i
n/\beta)],
\end{split}
\end{equation}
where we have used the property $\ell(-x)=-\ell(x)$. With the aid of
(\ref{eq:ell-Psi}) we can write the above result in terms of the
$\psi$ functions as follows
\begin{equation}
\begin{split}
\ln Z^{SUSY}_{\chi}(R,\beta)&=\sum_{n>0}\ln\frac{\psi(R-1-2\pi i
n/\beta)}{\psi(1-R-2\pi i
n/\beta)}+2\sum_{n>0}\frac{2\pi^2(R-1)n}{\beta}+\ell(-(R-1))\\
&=\sum_{n>0}\ln\frac{\psi(R-1-2\pi i n/\beta)}{\psi(1-R-2\pi i
n/\beta)}-\frac{\pi^2(R-1)}{3\beta}+\ell(-(R-1)),\label{eq:combineNoSq}
\end{split}
\end{equation}
where, following \cite{DiPietro:2014}, we have used zeta function
regularization in order to arrive at the result in the last line.

Note that our computation differs from that given in the appendix of
\cite{DiPietro:2014} in a few respects.  This affects the subleading
order, but not the leading order result, which was the focus of
\cite{DiPietro:2014}. First of all, in contrast with
\cite{DiPietro:2014}, we have assembled the KK contributions before
taking the high-temperature limit.  Secondly, while the 3d bosonic
partition functions in \cite{DiPietro:2014} and \cite{Hama:2011} are
identical, agreement of the 3d fermionic partition functions is more
subtle. The Dirac spectrum in (4.6) of \cite{Hama:2011} has two
pieces; if the contribution from the second term is summed over
\emph{after shifting} the related quantum number, and a relative
minus sign is introduced,
then (A.4) of \cite{DiPietro:2014} is recovered%
\footnote{Note that $l$, $q$, $n$, and $\sigma$ in \cite{Hama:2011}
correspond to $r_3$, $R$, $l$, and $n/r_1$ in \cite{DiPietro:2014},
respectively.}.
To avoid subtleties with the fermion reduction (such as the mixing
of the reduced fermions), our logic above is to reduce the bosons on
$S^1$ to obtain the 3d bosonic Lagrangian. Then, instead of reducing
the fermions, we simply appeal to the SUSY completion of the 3d
action of the KK bosons. The resulting 3d partition function for the
$n$-th KK modes is now that reported in the 3d localization
literature. This method of computing the 4d partition function is
equivalent to that of \cite{Closset:2014,Assel:2014}.

Comparing (\ref{eq:combineNoSq}) with (\ref{eq:GammaFV}),
(\ref{eq:M}), and (\ref{eq:ell-Psi}), now yields the relation
\begin{equation}
\mathcal{I}_{\chi}(R,\beta)=e^{\left(\frac{R^3}{6}-\frac{R^2}{2}+\frac{5R}{12}-\frac{1}{12}\right)\beta}
Z_{\chi}^{SUSY}(R,\beta).\label{eq:ZandIgenR}
\end{equation}
Since a chiral multiplet with R-charge $R$ has
\begin{equation}
\frac{R^3}{6}-\frac{R^2}{2}+\frac{5R}{12}-\frac{1}{12}=\frac{4}{27}(3c+a),
\end{equation}
we confirm the relation (\ref{eq:ZandIgen}) between the index and
the SUSY partition function for this case. Combining
(\ref{eq:ZandIgen}) with the high-temperature expansion of the index
in (\ref{eq:IchiExp}) then gives the expansion for $\ln Z^{SUSY}$
presented in (\ref{eq:conjZ}). Note in particular that the last term
on the RHS of Eq.~(\ref{eq:combineNoSq}) is the contribution of the
zero-modes. Our computation above therefore gives some understanding
for why the finite part of the 4d SUSY partition function reduces to
the 3d partition function upon taking the $\beta\rightarrow0$ limit;
this is because the $n\neq 0$ KK modes only contribute to the
$\mathcal{O}(1/\beta)$ term in $\ln Z^{SUSY}$, besides giving
transcendentally small corrections to it that are negligible in the
high-temperature limit.

\subsection{Free U(1) vector multiplet}

Our next case study is the theory of a single free U(1) vector
multiplet. The index of this theory is given by \cite{Dolan:2008}
\begin{equation}
\mathcal{I}_v(p,q)=(p;p)(q;q),
\end{equation}
where $(a;q)=\prod_{k=0}^\infty(1-aq^k)$ is the $q$-Pochhammer
symbol. We are, of course, mainly interested in the case of equal
fugacities, in which case
\begin{equation}
\mathcal{I}_v(\beta)=(q;q)^2.\label{eq:vectIndex}
\end{equation}
Note that $(q;q)$ is related to Dedekind's eta function via
\begin{equation}
\eta(\beta)=q^{1/24}(q;q).
\end{equation}

The high-temperature expansion may be obtained by invoking the
familiar SL(2,$\mathbb{Z}$) modular property
$\eta(-1/\tau)=\sqrt{-i\tau}\eta(\tau)$. We find that at high
temperatures $\eta\rightarrow
e^{-\pi^2/6\beta}\sqrt{\frac{2\pi}{\beta}}$, which leads to
\begin{equation}
\mathcal{I}_v(\beta)\rightarrow
e^{-\pi^2/3\beta}\left(\frac{2\pi}{\beta}\right)e^{\beta/12}.
\end{equation}
Taking the logarithm of this equation gives
\begin{equation}
\ln \mathcal{I}_v\sim
-\pi^2/3\beta-\ln\left(\frac{\beta}{2\pi}\right)+\beta/12,\label{eq:logIvExp}
\end{equation}
where again $\sim$ means correct to all orders in $\beta$, but
excluding non-analytic corrections of the type $e^{-1/\beta}$. This
is almost in full agreement with (\ref{eq:conjI}), since for a
single vector multiplet $c=1/8$ and $a=3/16$. In particular,
(\ref{eq:logIvExp}) confirms the conjecture in \cite{Ardehali:2015}
regarding the linear term. Also, terms of order $\beta^2$ or higher
are absent. To make the agreement with (\ref{eq:conjI}) complete,
however, we need to have $\ln Z_{3d}=0$. As explained in
\cite{DiPietro:2014} the dimensionally reduced theory of a vector
multiplet is quantum-mechanically ill-defined, and the logarithmic
term in (\ref{eq:logIvExp}) is signalling a problem; the 3d vector
multiplet has non-compact moduli, over which
the vacuum state can not be properly normalized (see also the
related discussion in \cite{Buican:2014}). There would not be
anything puzzling with our finding $\ln Z_{3d}=0$, however, if we
think of (\ref{eq:vectIndex}) as the index of a U(1) vector
multiplet with the zero-modes removed (see \cite{Benini:2013} for a
similar terminology in a 2d context).

Now consider the SUSY partition function of the same theory. It is
given in Eq.~(G.11) of \cite{Assel:2014} (with $r_G=1$)
\begin{equation}
Z^{SUSY}_v(\beta)=e^{\frac{i\pi}{2}\Psi(0,\tau,\tau)}(q;q)^2=e^{\frac{i\pi}{2}\Psi(0,\tau,\tau)}\mathcal{I}_v(\beta),\label{eq:ZvPsi}
\end{equation}
where $\Psi(w,\tau,\tau)$ is defined through
\begin{equation}
F(w,\tau,\tau)F(-w,\tau,\tau)=\frac{e^{i\pi\Psi(w,\tau,\tau)}}{\Gamma(w,\tau,\tau)\Gamma(-w,\tau,\tau)},\label{eq:PsiDef}
\end{equation}
with
\begin{equation}
F(w,\tau,\sigma)=\prod_{n_0\in\mathbb{Z}}\prod_{n_1,n_2\ge0}\frac{w+n_0+\frac{\tau+\sigma}{2}-\frac{\tau+\sigma}{2}-n_1\tau-n_2\sigma}{w+n_0+\frac{\tau+\sigma}{2}+\frac{\tau+\sigma}{2}+n_1\tau+n_2\sigma}.
\end{equation}
Combining the previous two equations, we arrive at
\begin{equation}
\begin{split}
F(w,\tau,\sigma)F(-w,\tau,\sigma)=\prod_{n_0\in\mathbb{Z}}\biggl[\;\prod_{n_1,n_2\ge0}&\frac{(-w-n_0-\frac{\tau+\sigma}{2})+\frac{\tau+\sigma}{2}+n_1\tau+n_2\sigma}{(w+n_0+\frac{\tau+\sigma}{2})+\frac{\tau+\sigma}{2}+n_1\tau+n_2\sigma}\\
&\times\frac{(w-n_0-\frac{\tau+\sigma}{2})+\frac{\tau+\sigma}{2}+n_1\tau+n_2\sigma}{(-w+n_0+\frac{\tau+\sigma}{2})+\frac{\tau+\sigma}{2}+n_1\tau+n_2\sigma}\biggr].
\end{split}
\end{equation}
The RHS can be written in terms of the function $s_b$ in
\cite{Hama:2011}
\begin{equation}
\begin{split}
F(w,\tau,\tau)F(-w,\tau,\tau)&=\prod_{n_0\in\mathbb{Z}}\left[s_{b=1}\left(-i+\frac{2\pi}{\beta}(n_0+w)\right)
s_{b=1}\left(-i+\frac{2\pi}{\beta}(n_0-w)\right)\right]\\
&=\prod_{n_0\in\mathbb{Z}}\left[s_{b=1}\left(i-\frac{2\pi}{\beta}(n_0+w)\right)s_{b=1}\left(i-\frac{2\pi}{\beta}(n_0-w)\right)\right]^{-1}.
\end{split}
\end{equation}
In the last step we have used $s_b(-x)=1/s_b(x)$. Comparing with
(\ref{eq:nthKKchi}) makes it now clear that
\begin{equation}
F(w,\tau,\tau)F(-w,\tau,\tau)=\frac{1}{Z^{SUSY}_{\chi}(R=w/\tau,\beta)Z^{SUSY}_{\chi}(R=-w/\tau,\beta)}.\label{eq:FFsimplified}
\end{equation}
The conversion factor that we derived between $Z^{SUSY}_{\chi}$ and
$\mathcal{I}_{\chi}$ in (\ref{eq:ZandIgenR}) leads therefore to the
correct function $\Psi$ mediating $Z^{SUSY}_{v}$ and
$\mathcal{I}_v$. Explicit calculation by combining
(\ref{eq:IwithGamma}), (\ref{eq:ZandIgenR}), (\ref{eq:PsiDef}) and
(\ref{eq:FFsimplified}) shows
\begin{equation}
\Psi(w,\tau,\tau)=2\frac{w^2}{\tau}+\frac{\tau}{3}.\label{eq:PsiExplicit}
\end{equation}
Plugging back into (\ref{eq:ZvPsi}) gives
\begin{equation}
Z^{SUSY}_v(\beta)=\eta(q)^2=e^{-\beta/12}\mathcal{I}_v(\beta),\label{eq:ZvEta}
\end{equation}
which confirms (\ref{eq:ZandIgen}) for the free vector case. Using
Eq. (\ref{eq:logIvExp}) we can then write down the following
high-temperature expansion
\begin{equation}
\ln Z^{SUSY}_v\sim
-\pi^2/3\beta-\ln\left(\frac{\beta}{2\pi}\right).\label{eq:logZvExp}
\end{equation}

\section{A Di~Pietro-Komargodski formula for large-$N$ toric quivers}\label{sec:4}

Toric quiver theories are a much-studied subset of supersymmetric
gauge theories whose field content can be efficiently summarized in
a quiver diagram. These are directed graphs with nodes representing
$\mathcal{N}=1$ vector multiplets and edges representing
$\mathcal{N}=1$ chiral multiplets. The nodes at the ends of an
edge represent vector multiplets under which the chiral multiplet represented by the
edge is charged. The direction of the edge encodes further
information about the representation of the gauge group according to
which the chiral multiplet transforms. The toric condition puts
further constraints on the theory, thereby guaranteeing some nice
properties such as existence of a non-trivial IR fixed point with a
holographic dual describable by ``toric geometry'' (see for instance
\cite{Franco:2006}). A canonical example is the $\mathcal{N}=4$ SYM
with SU($N$) gauge group, which can be represented by one node
(standing for the SU($N$) vector multiplet), and three directed
edges (standing for the three $\mathcal{N}=1$ chiral multiplets in
the adjoint) that both emanate from and end on that one node.

In this section we show that for the large-$N$ limit of toric quiver
gauge theories the relation (\ref{eq:DK}) is modified to
\begin{equation} \ln
\mathcal{I}^{N\rightarrow\infty}_{quiver}(\beta)\sim
\frac{\pi^2}{6\beta}\sum_{i=1}^{n_z}\frac{1}{r_i}+\frac{16\pi^2}{3\beta}\sum_{adj}(\delta
c_{adj}-\delta a_{adj})+\frac{n_z}{2}\ln(\beta/2\pi)+\ln
Y+\frac{4(3\delta c+\delta a)\beta}{27},\label{eq:conjLargeN}
\end{equation}
where $r_i$ are the R-charges of extremal BPS mesons in the quiver
\cite{Eager:2012hx,Agarwal:2013}, $n_z$ is the number of such mesons
(or the number of corresponding zigzag paths in the brane-tiling
picture \cite{Agarwal:2013}), $\delta c$ and $\delta a$ denote the
$\mathcal O(1)$ contributions to the full central charges (while
$\delta c_{adj}$ and $\delta a_{adj}$ denote only the contributions
from any chiral adjoint matter) and $\ln
Y=\frac{1}{2}\sum_{i=1}^{n_z}\ln r_i+\sum_{adj}\ell(R_{adj}-1)$. See
equations (\ref{eq:ell-Psi}) and (\ref{eq:psidef}) for the
definition of the function $\ell$. The way to determine $r_i$ and
$n_z$ for a given quiver is explained in Eq. (\ref{eq:ESTfact})
below.

With the aid of a conjecture in \cite{Ardehali:2014}, we can
write\footnote{In \cite{Ardehali:2014} the conjectured expression
was given for $a_0-b_0$, the difference of two coefficients
appearing in the high-temperature expansion of the single-trace
index. However, in all cases considered there $a_0-b_0=\sum1/r_i$,
so the conjecture can be alternatively stated as in
(\ref{eq:sumOneOverRi}).}
\begin{equation}
\sum_{i=1}^{n_z}\frac{1}{r_i}=\frac{3}{16\pi^3}\left(19\mathrm{vol}(SE)+\frac{1}{8}\mathrm{Riem}^2(SE)\right),\label{eq:sumOneOverRi}
\end{equation}
where $SE$ denotes the Sasaki-Einstein 5-manifold dual to the quiver
gauge theory. The above conjecture was motivated by the finding in
\cite{Eager:2010} that one can ``hear the shape of the dual
geometry'' in the asymptotics of the Hilbert series of mesonic
operators in the SCFT. We note that the leading high-temperature
behavior of the index of toric quivers is contained in the first two
terms of (\ref{eq:conjLargeN}). The first term, according to
(\ref{eq:sumOneOverRi}), is dictated by the geometry of the dual
internal manifold, while the second is given by the $\mathcal{O}(1)$
part of the contribution of adjoint matter to $c-a$.  The latter is
hence the only part of the finite-$N$ Di~Pietro-Komargodski formula
that escapes metamorphosis into ``geometry'' in the planar limit. In
addition, while at zero squashing both of these terms have the same
dependence on $\beta,$ as displayed in (\ref{eq:conjLargeNsq}) they
each have distinct dependence on the squashing parameter and can
therefore be distinguished.

As an illustrative example, let us consider the
$\mathcal{N}=4$ theory, and see how the conjecture (\ref{eq:sumOneOverRi})
works for this case. In this theory $n_z=3$ and%
\footnote{Note that $r=2/3$ is the R-charge of the trace of the
adjoint matter. This exemplifies the fact that the ``extremal BPS
mesons'' that play a role in (\ref{eq:conjLargeN}) are in general
mesons of the theory with U($N$) gauge group. The language of zigzag
paths \cite{Agarwal:2013} might therefore be preferable when
studying SU($N$) quivers.}
$r_{1,2,3}=2/3$. The conjecture (\ref{eq:sumOneOverRi}) reads
\begin{equation}
\frac{9}{2}=\sum_{i=1}^{3}\frac{1}{r_i}=\frac{3}{16\pi^3}\left(19\mathrm{vol}(S^5)+\frac{1}{8}\mathrm{Riem}^2(S^5)\right)=\frac{3}{16\pi^3}\left(19(\pi^3)+\frac{1}{8}(40\pi^3)\right),\label{eq:sumOneOverRiN=4}
\end{equation}
where we have used the geometrical data in Table~2 of
\cite{Eager:2010} to evaluate the RHS. Similar tests can be
successfully performed for all the SE$_5$ manifolds listed in
Table~2 of \cite{Eager:2010}.

\subsection{Derivation}

Our starting point for computing the large-$N$ index is the
following expression, valid when the nodes of the quiver have
SU($N$) gauge groups \cite{Gadde:2010en}
\begin{equation}
\ln\mathcal{I}^{N\rightarrow\infty}_{quiver}(q)=-\sum_{k=1}^{\infty}
\frac{\mathrm{tr}\,i(q^k)}{k}-\ln\prod_{k=1}^{\infty}{\mathrm{det}(1-i(q^k))}.\label{eq:largeNlog}
\end{equation}
The matrix $i$ has the single-letter index of the fields
transforming in the fundamental representation of the $j$-th node
and the anti-fundamental representation of the $k$-th node as its
$jk$ entry. On its diagonal it has the single-letter index of the
corresponding vector multiplets and the adjoint matter. The first
term on the RHS of (\ref{eq:largeNlog}) is the subtracted
contribution of the U(1)'s from the U($N$) answer given by the
second term. We neglect the first term until
Eq.~(\ref{eq:largeNquiverLog}) where it is re-introduced.

To obtain an expression for the second term on the RHS of
(\ref{eq:largeNlog}) we use \cite{Eager:2012hx}
\begin{equation}
(1-i(q))=\frac{\chi(q)}{(1-q)^2},\label{eq:EST}
\end{equation}
with the $n_v\times n_v$ matrix $\chi$ (where $n_v$ is the number of
nodes in the quiver) being a purely graph-theoretic object given by
\begin{equation}
\chi(q)=1-q^2-M_Q(q)+q^2 M_Q(q^{-1}).\label{eq:ESTmatrixDef}
\end{equation}
Here $M_Q(q)$ is the weighted adjacency matrix
\begin{equation}
M_Q(q)=\sum_e q^{R(e)}E_{h(e),t(e)},
\end{equation}
with $R(e)$ the R-charge of the edge $e$ in the quiver and $E_{v,w}$
is a matrix such that the $(v,w)$ entry is $1$ and all other entries
are zero.

Our following manipulations are made possible by the remarkable
factorization \cite{Eager:2012hx,Agarwal:2013}
\begin{equation}
\mathrm{det}\chi(t)=\prod_{i=1}^{n_z}(1-t^{r_i}).\label{eq:ESTfact}
\end{equation}
The above identity is proven for a subset of all toric quivers in
\cite{Eager:2012hx}, but is conjectured to be valid more generally
\cite{Agarwal:2013}. It allows an efficient rewriting of the index
of the quiver theories in terms of $(q^{r_i};q^{r_i})$.

The second term on the RHS of (\ref{eq:largeNlog}) can be written as
\begin{equation}
-\ln\prod_{k=1}^{\infty}{\mathrm{det}\left(\frac{\chi(q^k)}{(1-q^k)^2}\right)}=-\ln\prod_{k=1}^{\infty}
\left(\frac{1}{(1-q^k)^{2n_v}}\right){\mathrm{det}(\chi(q^k))},
\end{equation}
which with the aid of the $q$-Pochhammer symbol and
Eq.~(\ref{eq:vectIndex}) can be cast into
\begin{equation}
2n_v\ln(q;q)-\sum_{i=1}^{n_z}\ln(t^{r_i};t^{r_i})=n_v\ln\mathcal{I}_v(\beta)-\frac{1}{2}\sum_{i=1}^{n_z}\ln\mathcal{I}_v(\beta
r_i).\label{eq:U(N)index}
\end{equation}

Now we re-introduce the first term on the RHS of
(\ref{eq:largeNlog}), whose contribution from the vector multiplets
happens to kill the first term on the RHS of the above equation.
However, the contribution from the adjoint matter remains, so that
\begin{equation}
\ln\mathcal{I}^{N\rightarrow\infty}_{quiver}(\beta)=-\frac{1}{2}\sum_{i=1}^{n_z}\ln\mathcal{I}_v(\beta
r_i)-\sum_{adj} \ln\mathcal{I}_\chi
(R_{adj},\beta).\label{eq:largeNquiverLog}
\end{equation}
Using $\sum_{i=1}^{n_z}r_i=2n_v$, and also employing
(\ref{eq:IchiExp}) and (\ref{eq:logIvExp}), the high-temperature
expansion given in (\ref{eq:conjLargeN}) is obtained.

\subsection{The $\mathcal{N}=4$ theory as an example}

The $\mathcal{N}=4$ theory has one vector multiplet and three
adjoint chiral multiplets of R-charge $R=2/3$. Application of
Eq.~(\ref{eq:ESTfact}) gives for this case $n_z=3$ and
$r_{1,2,3}=2/3$. For the theory with SU($N$) gauge group we find
from (\ref{eq:largeNquiverLog}) that
\begin{equation}
\ln\mathcal{I}^{N\rightarrow\infty}_{\mathcal{N}=4}(\beta)=-\frac{3}{2}\ln\mathcal{I}_v(2\beta/3)-3
\ln\mathcal{I}_\chi (R=2/3,\beta)=-3\ln(q^{2/3};q^{2/3})-3
\ln\Gamma(2\tau/3,\tau,\tau).\label{eq:largeNquiverLogN=4}
\end{equation}
The asymptotic high-temperature expansion can be derived from the
expressions in the previous section to be
\begin{equation}
\ln\mathcal{I}^{N\rightarrow\infty}_{\mathcal{N}=4}(\beta)\sim\frac{5}{2}\frac{\pi^2}{6\beta}+\frac{3}{2}\ln\left(\frac{\beta}{2\pi}\right)+\left(\frac{3}{2}\ln\frac{2}{3}-3\ell(1/3)\right)-\frac{4}{27}\beta,
\end{equation}
which is exact up to non-analytic corrections of the type
$e^{-1/\beta}$.

Note that for the $\mathcal{N}=4$ theory with U($N$) gauge group we
would only need the second term in Eq. (\ref{eq:largeNlog}), whose
representation in (\ref{eq:U(N)index}) can be employed to obtain
\begin{equation}
\ln\mathcal{I}^{N\rightarrow\infty}_{\mathrm{U}(N)\
\mathcal{N}=4}(\beta)\sim\frac{1}{2}\frac{\pi^2}{6\beta}+\frac{1}{2}\ln\left(\frac{\beta}{2\pi}\right)+\frac{3}{2}\ln\frac{2}{3},
\end{equation}
with no $\mathcal{O}(\beta)$ term on the RHS, and also no appearance
of the $\ell$ function. Both of these features are shared by all
U($N$) quivers described by (\ref{eq:U(N)index}).

\section{Discussion}

In this note we have considered free chiral and U(1) vector
multiplets, and shown the robustness of the Di~Pietro-Komargodski
formula. When written as the high-temperature expansion of $\ln
Z^{SUSY}(\beta)$, it receives---aside from transcendentally small
contributions---only $\mathcal{O}(\ln\beta)$ and
$\mathcal{O}(\beta^0)$ corrections. It is tempting to speculate that
similar statements apply to general Lagrangian SCFTs. It may be
possible to investigate this robustness by defining a ``holomorphic
temperature'' and using holomorphy on $S^3\times S^1$
\cite{Festuccia:2011}. Another suspicion is that a yet more robust
version of the Di~Pietro-Komargodski formula may exist for SUSY
partition functions obtained by path-integration over
\emph{holomorphically normalized}, as opposed to canonically
normalized, gauge fields. To explore these possibilities, extending
the ``effective gauge coupling'' technique of \cite{Yaffe:1982}, and
``holomorphic gauge coupling'' technique of \cite{Arkani-Hamed:2000}
to the curved-space supersymmetric case may prove helpful.

The reader may ask why in our discussion of finite-$N$ theories we
have emphasized that the theories under study are free, while the
index of any Lagrangian theory is independent of the couplings and
can be evaluated easily even if the theory flows to an interacting
SCFT in the IR. The reason is that such interacting SCFTs can not be
constructed with only chiral and U(1) vector multiplets; one needs
asymptotic freedom in the Lagrangian, and therefore non-abelian
gauge fields. We have not studied non-abelian gauge theories in this
work because their index is significantly more difficult to analyze,
involving contour integrals that are hard to evaluate analytically
\cite{Dolan:2008}. We hope that the understanding gained in this
work eventually help analyzing the high-temperature behavior of
non-abelian gauge fields.

The subleading $\mathcal{O}(\ln\beta)$ term in (\ref{eq:conjZ}) and
(\ref{eq:conjI}) has some resemblance to the results of
\cite{Shapere:2008,Buican:2014}; it may be possible to make
immediate progress generalizing (\ref{eq:conjZ}) and
(\ref{eq:conjI}) for $\mathcal{N}=2$ SCFTs by their methods. It
would be very interesting if an analysis along those lines shows
that the coefficient of $\ln\beta$ in (\ref{eq:conjZ}) and
(\ref{eq:conjI}) depends---in contrast to what we claimed---not only
on the central charges, but also on some ``non-universal''
information, such as the dimensions of certain operators.

Another direction to study is examining the high-temperature
behavior of the index of non-Lagrangian SCFTs. The particular case
of $E_6$ SCFT is readily in analytical reach \cite{Gadde:2010}. In
fact, since the proof of Di~Pietro and Komargodski applies only to
Lagrangian theories \cite{DiPietro:2014}, it would be very
interesting to see if even the leading high-temperature behavior
pans out for the $E_6$ SCFT.

A somewhat different path to explore is that of large-$N$ gauge
theories. For toric quivers we presented in
Eq.~(\ref{eq:conjLargeN}) the explicit form of the modified
Di~Pietro-Komargodski formula, including its subleading corrections
to all orders in $\beta$. As Eqs.~(\ref{eq:conjLargeN}) and
(\ref{eq:sumOneOverRi}) show, understanding the high-temperature
behavior of the index of holographic quivers involves elements of
graph theory, geometry, and the theory of modular forms (or perhaps
a matrix generalization thereof, as Eq.~(\ref{eq:largeNlog})
suggests). It would be nice to have a more general understanding of
the connection between these elements beyond the toric case.

A topic we did not touch upon in our discussion of large-$N$ quivers
is that of the SUSY partition functions
$Z^{N\rightarrow\infty}_{quiver}(\beta)$ obtained by path
integration, in the planar limit. In analogy with the finite-$N$
case, we expect such partition functions to be proportional to
$\mathcal{I}^{N\rightarrow\infty}_{quiver}(\beta)$ with an
anomaly-dependent coefficient mediating the relation. We leave a
careful study of this problem to the future, and simply note that a
modification of the finite-$N$ version in (\ref{eq:ZandIgen}) may
have implications for the Casimir energy mismatch puzzle raised in
\cite{Cassani:2014}.

Non-holographic theories in the planar limit present another
playground in which to observe potential modifications of the
Di~Pietro-Komargodski formula, and explore its subleading
corrections. Let us discuss one example of this class, namely the
SQCD fixed point with $x=N_c/N_f=1/2$ in the Veneziano limit. The
index of this theory can be easily obtained from the expressions
given in \cite{Dolan:2008}. We find
\begin{equation}
\begin{split}
\ln\mathcal{I}^{N_c\rightarrow\infty}_{SQCD,x=1/2}(\beta)&=-N_f^2\ln
(q;q)+(N_f^2-1)\ln
(q^2;q^2)\\
&\sim\frac{N_f^2+1}{2}\left(\frac{\pi^2}{6\beta}\right)+\frac{1}{2}\ln\left(\frac{\beta}{2\pi}\right)
-\frac{N_f^2-1}{2}\ln
2+\frac{N_f^2-2}{24}\beta.\label{eq:SQCDonehalf}
\end{split}
\end{equation}
The coefficient of ${\pi^2}/{6\beta}$ in the second line is known
(as the $a_0$ coefficient in Table 2 of \cite{Ardehali:2014}) to be
$2N_c^2+1/2$ for general $x$. An application of the finite-$N$
Di~Pietro-Komargodski formula would give instead $32(c-a)=2N_c^2+2$,
which although correct at order $N_c^2$, differs at order one from
the actual value. On the other hand the coefficient of $\beta$ above
precisely matches with $4(3c+a)/27$ predicted by the finite-$N$
formula (\ref{eq:conjI}). Importantly, unlike for the holographic
quivers, this term includes the full central charges, and not just
their $\mathcal{O}(1)$ piece. This is related to the observations
made in \cite{Ardehali:2015} regarding the possibility of extracting
the full central charges from the large-$N$ index of $A_k$ SQCD
fixed points (see appendix \ref{app:B}).

Finally, modular properties of the SUSY partition functions
discussed above hint toward a general modular structure in four
dimensions. In section \ref{sec:2} we presented some 2d relations
that bear striking resemblance to those in four dimensions. The
resemblance is, however, far from perfect. For example, $Z_{PI}$ in
section \ref{sec:2} was modular invariant, but the four-dimensional
$Z^{SUSY}$ is apparently not. A deeper understanding of the
differences with the 2d modular structure may shed light on the
4d/2d relations
\cite{AGT,Wyllard:2009,Nekrasov:2010,Gadde:2011rry,Yamazaki:2012,Yamazaki:2012b}.

\subsection*{Note added:} Shortly after the first version of the
present paper appeared on arXiv, two important related developments
happened. The authors of \cite{Benjamin:2015} formulated a
sparseness condition for the elliptic genera of 2d CFTs with (2,2)
supersymmetry, partially addressing the problem mentioned in the
last sentence of section \ref{sec:2}. In \cite{Assel:2015s} the
supersymmetric Casimir energy is studied in great detail, and also
the regularization of SUSY partition functions on Hopf surfaces is
clarified.

\acknowledgments

A.A.A wishes to thank F.~Larsen for helpful conversations on 2d
CFTs, and A.~Gadde for a valuable comment regarding the index of a
chiral multiplet. P.S. would like to thank N.~Bobev for useful
discussions. We are also grateful to M.~Goykhman for communicating
with us unpublished results that inspired our discussion in appendix
\ref{app:B}. This work is part of the D-ITP consortium, a program of
the Netherlands Organisation for Scientific Research (NWO) that is
funded by the Dutch Ministry of Education, Culture and Science
(OCW), and is also supported in part by the US Department of Energy
under grant DE-SC0007859.


\appendix

\section{SCFTs with squashed three-sphere as spatial
manifold}\label{app:A}

In the main body, we have focused on the round $S^3\times S^1$. Here
we demonstrate that the results can be extended to the more general
case where the fugacities $p$ and $q$ are not necessarily equal.
The index with two fugacities is given in (\ref{eq:IndexTwofugDef}),
and the corresponding SUSY partition function is (see for instance
\cite{Aharony:2013}) the one computed on $S_b^3\times S^1$, with the
first factor representing a squashed 3-sphere with squashing
parameter $b$, which is related to the fugacities by $p=e^{2\pi i
\tau}=e^{-\beta b^{-1}}$ and $q=e^{2\pi i \sigma}=e^{-\beta b}$.

In the following we will need some mathematical notation that we now
introduce. An important role will be played below by the ``double
sine'' function $s_b$ in \cite{Hama:2011sq}, which can be
represented as
\begin{equation}
s_b(-ix)=\prod_{m,n\ge0}\frac{mb+nb^{-1}+\frac{Q}{2}-x}{mb+nb^{-1}+\frac{Q}{2}+x},\label{eq:sbDef}
\end{equation}
with $Q\equiv b+b^{-1}$. In parallel with (\ref{eq:ell-s}) we can
define $\ell_b(x)=\ln s_b (ix)$. To generalize (\ref{eq:ell-Psi}) we
then define the function $\psi_b$ through
\begin{equation}
\ell_b(x)=\ln\psi_b(-x)+\fft{i\pi
x^2}2-\frac{i\pi}{24}(b^2+b^{-2}).\label{eq:ell-PsiSq}
\end{equation}
$\psi_b$ is related to the ``non-compact quantum dilogarithm''
function $e_b$ in \cite{Faddeev:2001} via $\psi_b(x)=e_b(-ix)$.
Using Eq.~(15) in \cite{Faddeev:2001} we can express $\psi_b(x)$ for
$|\mathrm{Re}\,x|<Q/2$ by
\begin{equation}
\ln\psi_b(x)=\int_{-\infty}^{+\infty}\frac{\mathrm{d}t}{4t}\frac{e^{-2xt}}{\sinh(bt)\sinh(t/b)},\label{eq:psiSqDef}
\end{equation}
where the singularity at $t=0$ is put below the contour of
integration. Explicit evaluation of the above contour integral for
$b=1$, gives $\psi$ as written in (\ref{eq:psidef}). In other words,
$\psi_{b=1}$ equals the $\psi$ function in the main text.
Importantly for our computations below, Eq.~(21) in
\cite{Faddeev:2001} can be used to write
\begin{equation}
\psi_b(x)=\frac{(e^{-2\pi i x b+i\pi Qb};e^{2\pi i b^2})}{(e^{-2\pi
ixb^{-1}-i\pi Qb^{-1}};e^{-2\pi i
b^{-2}})}=\prod_{k\ge0}\frac{1-e^{-2\pi i xb+i\pi Qb+2\pi i k
b^2}}{1-e^{-2\pi i xb^{-1}-i\pi Qb^{-1}-2\pi i k
b^{-2}}}.\label{eq:psiSqPoch}
\end{equation}

Note that when $x$ has a large negative imaginary part of order
$1/\beta$, the expression (\ref{eq:psiSqPoch}) shows that
$\ln\psi_b(x)\sim 1$, where $\sim$ means equality to all orders in
$\beta$ but excluding non-analytic corrections of the type
$e^{-1/\beta}$.

\subsection{Free chiral multiplet}\label{app:1}

Consider a chiral multiplet of R-charge $R$. The index is given by
\cite{Dolan:2008}
\begin{equation}
\mathcal{I}_\chi(R,\beta,b)=\Gamma(z,\tau,\sigma),
\end{equation}
with $z=R\frac{Q}{2}\frac{i\beta}{2\pi}$,
$\tau=\frac{i\beta}{2\pi}b^{-1}$, and $\sigma=\frac{i\beta}{2\pi}b$.
We will show below that (\ref{eq:GammaFV}) can be generalized to
\begin{equation}
\Gamma(z,\tau,\sigma)=\frac{e^{-i\pi
M(z;\tau,\sigma)}}{\psi_b(-(R-1)\frac{Q}{2})}\prod_{n=1}^{\infty}\frac{\psi_b(-\frac{2\pi
in}{\beta}+(R-1)\frac{Q}{2})}{\psi_b(-\frac{2\pi
in}{\beta}-(R-1)\frac{Q}{2})},\label{eq:GammaFVSq}
\end{equation}
where
\begin{equation}
\begin{split}
-i\pi M(z;\tau,\sigma)=&-\frac{\pi^2
(R-1)}{3\beta}\frac{Q}{2}-\left[\frac{i\pi}{2}\left((R-1)\frac{Q}{2}\right)^2-\frac{i\pi}{24}(b^2+b^{-2})\right]\\
&+\beta\left(-\frac{(R-1)}{48}(b+b^{-1}+b^3+b^{-3})+\frac{(R-1)^3}{48}(b+b^{-1})^3\right).\label{eq:MSq}
\end{split}
\end{equation}
Then an argument similar to that in the main text gives the
high-temperature expansion
\begin{equation}
\begin{split}
\ln\mathcal{I}_\chi(R,\beta,b)\sim &-\frac{\pi^2
(R-1)}{3\beta}\left(\frac{b+b^{-1}}{2}\right)+\ell_b\left(-(R-1)\frac{b+b^{-1}}{2}\right)\\
&+\beta\left(-\frac{(R-1)}{48}\left(b+\frac{1}{b}+b^3+\frac{1}{b^3}\right)+\frac{(R-1)^3}{48}\left(b+\frac{1}{b}\right)^3\right).\label{eq:logIsquashed}
\end{split}
\end{equation}
In particular, the index reduces to the 3d partition function,
$\exp(\ell_b(-(R-1)\frac{b+b^{-1}}{2}))=s_b(-i(R-1)\frac{b+b^{-1}}{2})$,
in the limit $\beta\rightarrow0$, after its $\mathcal{O}(1/\beta)$
divergent exponent is removed.

The expansion (\ref{eq:logIsquashed}) can be written in terms of the
central charges as (note that for chiral multiplets $2a-c=0$)
\begin{equation}
\begin{split}
\ln \mathcal{I}(\beta,b)\sim&
\frac{16\pi^2(c-a)}{3\beta}(\frac{b+b^{-1}}{2})-4(2a-c)\ln(\beta/2\pi)+\ln
Z_{3d}\\
&+\beta\left(\frac{2}{27}(b+b^{-1})^3(3c-2a)+\frac{2}{3}(b+b^{-1})(a-c)\right).\label{eq:logIsquashedCA}
\end{split}
\end{equation}
In section \ref{app:2} we will see that the above form applies (as
in the main text, up to the $\ln Z_{3d}$ term) to a U(1) vector
multiplet as well. Importantly, both for a free chiral and a free
U(1) vector multiplet, the linear term in the high-temperature
expansion of $\ln \mathcal{I}(\beta,b)$ encodes enough information
to allow extracting $a$ and $c$ separately from the index. As
demonstrated in appendix \ref{app:B}, this observation is crucial
for making contact with the prescriptions in \cite{Ardehali:2015}.

The calculation of the SUSY partition function (highlighted below)
goes through very similarly to that in the main text, and this time
yields
\begin{equation}
\ln
Z_\chi^{SUSY}(R,\beta,b)=\beta\left[\frac{(R-1)}{48}\left(b+\frac{1}{b}+b^3+\frac{1}{b^3}\right)-\frac{(R-1)^3}{48}\left(b+\frac{1}{b}\right)^3\right]+\ln\mathcal{I}_\chi(R,\beta,b),\label{eq:ZandIgenSqR}
\end{equation}
which can be written alternatively in terms of the central charges
of the chiral multiplet as
\begin{equation}
\boxed{\mathcal{I}(\beta,b)=\exp{\left[\beta\left(\frac{2}{27}(b+b^{-1})^3(3c-2a)+\frac{2}{3}(b+b^{-1})(a-c)\right)\right]}
\>Z^{SUSY}(\beta,b).\label{eq:ZandIgenSqca}}
\end{equation}
This agrees with the prefactors proposed in
\cite{Closset:2014,Assel:2014}, \emph{except for the absence of the
$\mathcal{O}(1/\beta)$ term} in the exponent.

Combining (\ref{eq:ZandIgenSqR}) and (\ref{eq:logIsquashed}) we
obtain the high-temperature expansion
\begin{equation}
\ln Z^{SUSY}_\chi(R,\beta,b)\sim -\frac{\pi^2
(R-1)}{3\beta}\left(\frac{b+b^{-1}}{2}\right)+\ell_b\left(-(R-1)\frac{b+b^{-1}}{2}\right),\label{eq:logZsquashed}
\end{equation}
or in terms of the central charges
\begin{equation} \boxed{\ln Z^{SUSY}(\beta,b)\sim
\frac{16\pi^2(c-a)}{3\beta}(\frac{b+b^{-1}}{2})-4(2a-c)\ln(\beta/2\pi)+\ln
Z_{3d},\label{eq:logZsquashedCA}}
\end{equation}
with no $\mathcal{O}(\beta)$ term on the RHS. Again, we will see in
section \ref{app:2} that (aside from the $\ln Z_{3d}$ term) the
above expansion applies to a U(1) vector multiplet as well.

We now turn to the proof of (\ref{eq:GammaFVSq}) by starting with
the modular property of the Gamma function (\ref{eq:modprop}).  We
rewrite this expression in terms of $R,b,\beta,Q$, expand using
(\ref{eq:ellGamma}), and manipulate as follows
\begin{equation}
\begin{split}
\Gamma(z,\tau,\sigma)=e^{-i\pi
M}&\frac{\Gamma(\frac{RQb}{2},\frac{2\pi
ib}{\beta},b^2)}{\Gamma(\frac{RQb^{-1}}{2}-b^{-2},\frac{2\pi
ib^{-1}}{\beta},-b^{-2})}\\
=e^{-i\pi M}&\prod_{n,k=0}^{\infty}\frac{1-e^{2\pi
i((n+1)(\frac{2\pi ib}{\beta})+(k+1)b^2-\frac{RQb}{2})}}{1-e^{2\pi
i(n(\frac{2\pi ib}{\beta})+kb^2+\frac{RQb}{2})}} \frac{1-e^{2\pi
i(n(\frac{2\pi
ib^{-1}}{\beta})-kb^{-2}+\frac{RQb^{-1}}{2}-b^{-2})}}{1-e^{2\pi
i((n+1)(\frac{2\pi
ib^{-1}}{\beta})-(k+1)b^{-2}-\frac{RQb^{-1}}{2}+b^{-2})}}\\
=e^{-i\pi
M}&\prod_{n,k=0}^{\infty}\frac{1-e^{-\frac{4\pi^2}{\beta}(n+1)b-i\pi
(R-1)Qb+i\pi Qb+2\pi
ikb^2}}{1-e^{-\frac{4\pi^2}{\beta}\frac{(n+1)}{b}-i\pi
(R-1)\frac{Q}{b}-i\pi \frac{Q}{b}-\frac{2\pi ik}{b^{2}}}}
\frac{1-e^{-\frac{4\pi^2}{\beta}\frac{n}{b}+i\pi
(R-1)\frac{Q}{b}-i\pi \frac{Q}{b}-\frac{2\pi
ik}{b^{2}}}}{1-e^{-\frac{4\pi^2}{\beta}nb+i\pi
(R-1)Qb+i\pi Qb+2\pi ikb^2}}\\
=e^{-i\pi M}&\prod_{k\ge0}\frac{1-e^{i\pi (R-1)Qb^{-1}-i\pi
Qb^{-1}-2\pi ikb^{-2}}}{1-e^{i\pi
(R-1)Qb+i\pi Qb+2\pi ikb^2}}\\
&\prod_{n>0,k\ge0}\frac{1-e^{-\frac{4\pi^2}{\beta}nb-i\pi
(R-1)Qb+i\pi Qb+2\pi ikb^2}}{1-e^{-\frac{4\pi^2}{\beta}nb^{-1}-i\pi
(R-1)Qb^{-1}-i\pi Qb^{-1}-2\pi ikb^{-2}}}\\
&\prod_{n>0,k\ge0}\frac{1-e^{-\frac{4\pi^2}{\beta}nb^{-1}+i\pi
(R-1)Qb^{-1}-i\pi Qb^{-1}-2\pi
ikb^{-2}}}{1-e^{-\frac{4\pi^2}{\beta}nb+i\pi (R-1)Qb+i\pi Qb+2\pi
ikb^2}}.
\end{split}
\end{equation}
Our claim in Eq.~(\ref{eq:GammaFVSq}) then follows from using the
expression (\ref{eq:psiSqPoch}) for the function $\psi_b$.

The computation of the SUSY partition function starts, as in
(\ref{eq:nthKKchi}), with the contribution of the $n$-th KK mode
\cite{Hama:2011sq}
\begin{equation}
\ln Z_\chi^{(n)}(R,\beta,b)=\ln
s_{b}\left(-i(R-1)\frac{Q}{2}-\frac{2\pi
n}{\beta}\right)=\ell_b\left(-(R-1)\frac{Q}{2}+\frac{2\pi i
n}{\beta}\right).\label{eq:nthKKchiSq}
\end{equation}
Then following a similar line of argument as in subsection
\ref{subsec:freeChi}, and using the relation (\ref{eq:ell-PsiSq})
between $\ell_b$ and $\psi_b$, we have
\begin{equation}
\begin{split}
\ln\!\!
Z^{SUSY}_{\chi}(R,\beta,b)&=\sum_{n>0}\ln\frac{\psi_b((R-1)\frac{Q}{2}-2\pi
i n/\beta)}{\psi_b((1-R)\frac{Q}{2}-2\pi i
n/\beta)}+2\sum_{n>0}\frac{2\pi^2(R-1)n}{\beta}\frac{Q}{2}+\ell_b\left(\!-(R-1)\frac{Q}{2}\right)\\
&=\sum_{n>0}\ln\frac{\psi_b((R-1)\frac{Q}{2}-2\pi i
n/\beta)}{\psi_b((1-R)\frac{Q}{2}-2\pi i
n/\beta)}-\frac{\pi^2(R-1)}{3\beta}\frac{Q}{2}+\ell_b\left(\!-(R-1)\frac{Q}{2}\right).\label{eq:combineWithSq}
\end{split}
\end{equation}
This combined with (\ref{eq:GammaFVSq}) proves
(\ref{eq:ZandIgenSqR}).

\subsection{Free U(1) vector multiplet}\label{app:2}

The superconformal index of a single free U(1) vector multiplet is
\cite{Dolan:2008}
\begin{equation}
\mathcal{I}_v(p,q)=(p;p)(q;q).
\end{equation}
Now, following the same line of argument that led to
(\ref{eq:logIvExp}) we arrive at
\begin{equation}
\mathcal{I}_v(p,q)\rightarrow
e^{-\pi^2(b+b^{-1})/6\beta}\left(\frac{2\pi}{\beta}\right)e^{(b+b^{-1})\beta/24},
\end{equation}
or
\begin{equation}
\ln \mathcal{I}_v(\beta,b)\sim
-\fft{\pi^2(b+b^{-1})}{6\beta}-\ln\left(\frac{\beta}{2\pi}\right)+\fft{\beta(b+b^{-1})}{24}.\label{eq:logIvTwofugExp}
\end{equation}

To study $Z^{SUSY}_v(p,q)$, we need $\Psi(w,\tau,\sigma)$. A
computation similar to that which led to (\ref{eq:PsiExplicit}), but
now employing (\ref{eq:ZandIgenSqR}), gives
\begin{equation}
\Psi(w,\tau,\sigma)=w^2\left(\frac{1}{\tau}+\frac{1}{\sigma}\right)+\frac{1}{6}(\tau+\sigma).\label{eq:PsiSq}
\end{equation}
For $w=0$, when combined with (G.11) of \cite{Assel:2014}, this
implies
\begin{equation}
Z^{SUSY}_v(p,q)=p^{1/24}q^{1/24}\mathcal{I}_v(p,q)=\eta(p)\eta(q),\label{eq:ZvSq}
\end{equation}
in accord with (\ref{eq:ZandIgenSqca}). The high-temperature
expansion is simply
\begin{equation}
\ln Z_v(\beta,b)\sim
-\fft{\pi^2(b+b^{-1})}{6\beta}-\ln\left(\frac{\beta}{2\pi}\right).\label{eq:logZvTwofugExp}
\end{equation}

Note that Eqs.~(\ref{eq:ZandIgenSqR}) and (\ref{eq:PsiSq}), when
combined with the localization results of \cite{Assel:2014}, yield
the relation (\ref{eq:ZandIgenSqca}) between $\mathcal{I}$ and
$Z^{SUSY}$ even in presence of interactions and non-abelian gauge
fields.

\subsection{Toric quivers in the planar limit}\label{subsec:toricApp}

To generalize our results in section \ref{sec:4} to the index with
two fugacities, we need the following more general form of
Eq.~(\ref{eq:EST}):
\begin{equation}
(1-i(p,q))=\frac{\chi(t)}{(1-p)(1-q)},\label{eq:ESTgen}
\end{equation}
where $t=\sqrt{pq}$. Eq.~(\ref{eq:ESTfact}) remains unchanged. The
rest of the computation in section \ref{sec:4} goes through with
little change, and one arrives at
\begin{equation}
\ln\mathcal{I}^{N\rightarrow\infty}_{quiver}(t,y)=-\frac{1}{2}\sum_{i=1}^{n_z}\ln\mathcal{I}_v(t_v=
t^{r_i},y_v=1)-\sum_{adj} \ln\mathcal{I}_\chi
(R=R_{adj},t_\chi=t,y_\chi=y),\label{eq:largeNquiverLogSq}
\end{equation}
where $y=\sqrt{p/q}$.

We can now use the results in appendices \ref{app:1} and \ref{app:2}
to write down the asymptotic high-temperature expansion of the
large-$N$ index in (\ref{eq:largeNquiverLogSq}). A simple
calculation using (\ref{eq:logIsquashed}) and
(\ref{eq:logIvTwofugExp}) shows
\begin{equation}
\boxed{\begin{aligned} \ln
\mathcal{I}^{N\rightarrow\infty}_{quiver}(\beta,b)\sim
&\,\frac{\pi^2}{6\beta(\frac{b+b^{-1}}{2})
}\sum_{i=1}^{n_z}\frac{1}{r_i}+\frac{16\pi^2
(\frac{b+b^{-1}}{2})}{3\beta}\sum_{adj}(\delta c_{adj}-\delta
a_{adj})+\frac{n_z}{2}\ln(\beta/2\pi)+\ln
Y_b \\
&\,+\beta\left(\frac{2}{27}(b+b^{-1})^3(3\delta c-2\delta
a)+\frac{2}{3}(b+b^{-1})(\delta a-\delta
c)\right),\label{eq:conjLargeNsq}
\end{aligned}}
\end{equation}
where the notation is similar to that in (\ref{eq:conjLargeN}),
except for $\ln Y_b=\frac{1}{2}\sum_{i=1}^{n_z}\ln (
r_i(\frac{b+b^{-1}}{2}))+\sum_{adj}\ell_b((R_{adj}-1)(\frac{b+b^{-1}}{2}))$.
See Eqs.~(\ref{eq:ell-PsiSq}) and (\ref{eq:psiSqDef}) for the
definition of the function $\ell_b$.

\section{Single-trace index and the central charges}\label{app:B}

The single-trace index is defined as the plethystic log
\cite{Benvenuti:2006} of the index (\ref{eq:IndexTwofugDef})
\begin{equation}
I_{s.t.}(\beta,b)\equiv\sum_{n=1}^{\infty}\frac{\mu(n)}{n}\ln
\mathcal{I}(n\beta,b),\label{eq:plLogDef}
\end{equation}
where $\mu(n)$ is the M\"{o}bius function. The adjective
``single-trace'' is particularly appropriate for theories that admit
a planar limit in which single-trace operators are weakly
interacting. For such cases if in the definition of the index in
(\ref{eq:IndexTwofugDef}) one restricts the trace to the
``single-trace states'' in the Hilbert space, one obtains the
single-trace index as defined above. In AdS/CFT, the weakly
interacting mesons of the SCFT at large 't~Hooft coupling map to the
KK supergravity modes in the bulk. Therefore the single-trace index
is quite natural from the bulk point of view.

In \cite{Ardehali:2015}, building on the holographic results of
\cite{Beccaria:2014}, prescriptions were proposed for extracting the
central charges $a$ and $c$ from the single-trace index of an SCFT.
It was observed that for holographic theories the prescriptions
reproduce the $\mathcal{O}(1)$ piece of the central charges, denoted
by $\delta a$ and $\delta c$, while for $A_k$ SQCD fixed points at
the Veneziano limit and for finite-$N$ theories they give the full
central charges and not just their $\mathcal{O}(1)$ piece. It was
also suspected that there may be a relation between those
prescriptions and the Di~Pietro-Komargodski formula (\ref{eq:DK}).

In this appendix we show that the proposals in \cite{Ardehali:2015}
probe in fact only the $\mathcal{O}(\beta)$ term in the
high-temperature expansion of the indices (\ref{eq:logIsquashedCA})
and (\ref{eq:conjLargeNsq}), and have nothing to do with their
leading $\mathcal{O}(1/\beta)$ behavior. This also explains the
applicability of the formulas in \cite{Ardehali:2015} to large-$N$
theories: while the leading $\mathcal{O}(1/\beta)$ behavior of the
large-$N$ indices can be drastically different from the finite-$N$
proposal of Di~Pietro and Komargodski, the $\mathcal{O}(\beta)$ term
is either completely inherited from the finite-$N$ theory (as in the
case of $A_k$ SQCD fixed points, an example of which appears in
Eq.~(\ref{eq:SQCDonehalf})), or at least its $\mathcal{O}(1)$ piece
survives (as in the case of holographic quivers described in
Eq.~(\ref{eq:conjLargeNsq})).

To simplify comparison with \cite{Ardehali:2015} we start by the
expansion of the single-trace index $I_{s.t.}$, first around $y=1$
and then around $t_t=1$ (see Sec.~IV of \cite{Ardehali:2014})
\begin{eqnarray}\label{eq:stIndexExpanded}
I_{s.t.}&=&\left(\fft{a_0}{t_t-1}+a_1+a_2(t_t-1)+\cdots\right)\nn\\
&&+(y-1)^2\left(\fft{b_0}{(t_t-1)^3}+\fft{b_1}{(t_t-1)^2}+\fft{b_2}{t_t-1}+\cdots\right)\nn\\
&&+\cdots,
\end{eqnarray}
where $t_t=1/t=e^{\beta Q/2}$ is the $t$-variable defined in
\cite{Ardehali:2015,Ardehali:2014} and
$y=e^{-\beta\left(\frac{b-1/b}{2}\right)}$. Then the prescriptions
in \cite{Ardehali:2015} read
\begin{eqnarray}\label{eq:cnaIndexExpanded}
\hat
a&=&\fft{9(a_0-b_0)}{32(t_t-1)^2}-\fft{3(a_0+12a_2)-9(b_0-b_1+b_2)}{32}+\cdots,\nn\\
\hat
c&=&-\fft{3(a_0-b_0)}{32(t_t-1)^2}-\fft{2(a_0+12a_2)+3(b_0-b_1+b_2)}{32}+\cdots,
\end{eqnarray}
where the functions $\hat a$ and $\hat c$ have a second-order pole
in the high-temperature ($t_t\rightarrow 1$) limit and their finite
piece in this limit gives the central charges (or in the case of
holographic quivers, their $\mathcal{O}(1)$ pieces).

\subsection{Finite-$N$ theories}\label{app:B1}

We start with the high-temperature expansion in
(\ref{eq:logIsquashedCA}), and take its plethystic log as defined in
Eq.~(\ref{eq:plLogDef}). The following sums are needed in the
process \cite{Nanxian:2010}
\begin{equation}\label{eq:muSums}
\begin{split}
\sum\frac{\mu(n)}{n^2}&=\frac{1}{\zeta(2)}=\frac{6}{\pi^2},\\
\sum\frac{\mu(n)}{n}&=0,\\
\sum\frac{\mu(n)\ln n}{n}&\rightarrow\frac{\zeta'(1)}{\zeta(1)^2}=-1,\\
\sum \mu(n)&\rightarrow\frac{1}{\zeta(0)}=-2.
\end{split}
\end{equation}
The result for the high-temperature expansion of the single-trace
index is
\begin{equation}
\begin{split}
I_{s.t.} \sim&
\frac{32(c-a)}{2\beta}\left(b+b^{-1}\right)+4(2a-c)\\
&-\beta\left(\frac{4}{27}(b+b^{-1})^3(3c-2a)+\frac{4}{3}(b+b^{-1})(a-c)\right).\label{eq:StExpanded}
\end{split}
\end{equation}

We now wish to demonstrate that the leading $\mathcal{O}(1/\beta)$
behavior of the single-trace index, or its subleading
$\mathcal{O}(\beta^0)$ term, do not play a role in evaluating the
central charges via (\ref{eq:cnaIndexExpanded}). To illustrate this,
we will assume the following form for the single-trace index
\begin{equation}
I_{s.t.}\sim
\frac{G}{2\beta}\left(b+b^{-1}\right)+C-\beta\left(\frac{4}{27}(b+b^{-1})^3(3c-2a)+\frac{4}{3}(b+b^{-1})(a-c)\right).\label{eq:StExpandedG}
\end{equation}
Using this, we will demonstrate that the prescriptions
(\ref{eq:cnaIndexExpanded}) are independent of the coefficients
$G,C.$

To compare (\ref{eq:StExpandedG}) with the expansion
(\ref{eq:stIndexExpanded}), we need the dictionary
\begin{equation}
b+b^{-1}=\frac{2\ln t_t}{\sqrt{\ln(t_ty)\ln(t_t/y)}},\quad
\beta=\sqrt{\ln(t_ty)\ln(t_t/y)}.
\end{equation}
Substituting the above relations in (\ref{eq:StExpandedG}),
expanding first around $y=1$ and then around $t_t=1$, and finally
comparing with (\ref{eq:stIndexExpanded}), gives
\begin{alignat}{3}\label{eq:Istcoeffs}
a_0&=G,\qquad &&a_1=\frac{1}{2}G+C ,\qquad &&a_2=-\frac{8}{27}a - \frac{8}{9}c - \frac{1}{12}G, \nonumber \\
b_0&=G,\qquad &&b_1=\frac{3}{2}G,\qquad &&b_2=\frac{64}{27}a -
\frac{32}{9}c + \frac{1}{2}G.
\end{alignat}
Plugging the above values in (\ref{eq:cnaIndexExpanded}) yields
\begin{eqnarray}
\hat a=a+\cdots,\quad \hat c=c+\cdots,
\end{eqnarray}
where the ellipses denote terms that vanish at $t_t=1$. Note that
the $G$-dependence of the coefficients (\ref{eq:Istcoeffs})  drops
out when evaluating (\ref{eq:cnaIndexExpanded}). Moreover, the only
$C$-dependent coefficient in (\ref{eq:Istcoeffs}) is $a_1$, which
does not show up in (\ref{eq:cnaIndexExpanded}).

\subsection{$A_k$ SQCD fixed points in the Veneziano limit}

A natural generalization of standard SQCD is achieved by adding a single
adjoint chiral multiplet $X$ to SQCD, and turning on a simple superpotential $\mathrm{Tr} X^{k+1}.$
This leads to the so-called $A_k$ SQCD theories \cite{Kutasov:1995np}.

We have not been able to compute the high-temperature expansion of
the index of these theories when $b\neq1$. In fact even for $b=1$,
SQCD (corresponding to $k=1$) with $x=1/2$ has been the only example
whose high-temperature expansion we have completely evaluated; it is
shown in (\ref{eq:SQCDonehalf}). Nevertheless, since the
single-trace index of these theories were computed explicitly in
\cite{Ardehali:2014}, we can go in the reverse direction to that in
the previous subsection, and use the high-temperature expansion of
$I_{s.t.}(\beta,b)$ to gain information on the expansion of $\ln
\mathcal{I}(\beta,b)$. This is, in fact, how the linear term in
(\ref{eq:conjI}) was conjectured in \cite{Ardehali:2015}. Note that,
as stated in \cite{Ardehali:2015}, going from the high-temperature
expansion of the single-trace index to that of $\ln
\mathcal{I}(\beta,b)$, one can not reproduce the $\mathcal{O}(\beta^0)$ term in the
latter. However, the other claim in \cite{Ardehali:2015}, that even the $\mathcal{O}(\ln\beta)$ term of $\ln
\mathcal{I}(\beta,b)$ can not be obtained from the high-temperature expansion of the single-trace index, is not quite correct. Here we see, from the third line
of (\ref{eq:muSums}), that the coefficeint of $\ln\beta$ in the expansion of $\ln
\mathcal{I}(\beta,b)$ is simply given by minus the $\mathcal{O}(\beta^0)$ term in the expansion of $I_{s.t.}(\beta,b)$.
For instance, by negating the second term on the RHS of relation (\ref{eq:StExpanded}) one obtains the coefficient of the log term on the RHS of (\ref{eq:logIsquashedCA}).

We leave out the details, and only report an ansatz which is
confirmed by the type of analysis mentioned in the previous
paragraph
\begin{equation}
\begin{split}
\ln\mathcal{I}^{N_c\rightarrow\infty}_{A_k\
SQCD}(\beta,b)\sim&\frac{2k^3+3k^2-1}{4k(1+k)}\left(\frac{\pi^2}{6\beta(\frac{b+b^{-1}}{2})}\right)+\frac{16kN_c^2-8k^2+8k}{4k(1+k)}\left(\frac{\pi^2(\frac{b+b^{-1}}{2})}{6\beta}\right)\\
&+\frac{1}{2}\ln (\frac{\beta}{2\pi})+\mathcal{O}(\beta^0)
+\beta\left(\frac{2}{27}(b+b^{-1})^3(3c-2a)+\frac{2}{3}(b+b^{-1})(a-c)\right).\label{eq:AkSQCDconj}
\end{split}
\end{equation}
In particular, the leading $\mathcal{O}(1/\beta)$ term is different
from the finite-$N$ Di~Pietro-Komargodski formula in two important
respects: $i)$ it is not entirely determined by the central charges,
and $ii)$ the dependence on $b+b^{-1}$ is slightly more complicated
than in the finite-$N$ version, (\ref{eq:logIsquashedCA}). While at
finite $N$ the leading term depends on $b+b^{-1}$ only through
$\frac{b+b^{-1}}{\beta}$, the large-$N$ version has another
$\mathcal{O}(1/\beta)$ term which is proportional to
$\frac{1}{(b+b^{-1})\beta}$. Similar observations could be made
equally as well for the high-temperature expansion of the index of
large-$N$ toric quivers, presented in (\ref{eq:conjLargeNsq}). In
both cases, the term proportional to $\frac{1}{(b+b^{-1})\beta}$,
which would be absent at finite $N$, is responsible for the poles
showing up in the prescriptions (\ref{eq:cnaIndexExpanded}). This
provides some insight into the divergences that were encountered in
\cite{Ardehali:2015,Ardehali:2014} when extracting the central
charges of large-$N$ theories.

In the next subsection we focus on large-$N$ quivers and demonstrate
the aforementioned connection between the poles in
(\ref{eq:cnaIndexExpanded}) and the term proportional to
$\frac{1}{(b+b^{-1})\beta}$ in the expansion of $\ln
\mathcal{I}(\beta,b)$. A completely similar analysis can be applied
to the $A_k$ SQCD theories.

\subsection{Toric quivers in the planar limit}

As in appendix \ref{app:B1}, we start with the high-temperature
expansion of the index, take its plethystic log to arrive at the
expansion of the single-trace index, rewrite it in terms of $t_t,y$,
and extract its coefficients to plug in the prescriptions
(\ref{eq:cnaIndexExpanded}). Taking the plethystic log of
(\ref{eq:conjLargeNsq}), we arrive at
\begin{equation}
\begin{split}
I_{s.t.\ quiver}^{N\rightarrow\infty}\sim&\,
\frac{2}{\beta\left(b+b^{-1}\right)}\sum_{i=1}^{n_z}\frac{1}{r_i}+\frac{32\left(b+b^{-1}\right)}{2\beta}\sum_{adj}(\delta c_{adj}-\delta a_{adj})-\frac{n_z}{2}\\
&\,-\beta\left(\frac{4}{27}(b+b^{-1})^3\left(3\delta c-2\delta
a\right)+\frac{4}{3}(b+b^{-1})\left(\delta a-\delta
c\right)\right).\label{eq:StExpandedQuiver}
\end{split}
\end{equation}

As in subsection \ref{app:B1}, we now assume the following expansion
instead of (\ref{eq:StExpandedQuiver})
\begin{equation}
\begin{split}
I_{s.t.\ quiver}^{N\rightarrow\infty}\sim&\,
\frac{2H}{\beta\left(b+b^{-1}\right)}+\frac{G\left(b+b^{-1}\right)}{2\beta}+C\\
&\,-\beta\left(\frac{4}{27}(b+b^{-1})^3\left(3\delta c-2\delta
a\right)+\frac{4}{3}(b+b^{-1})\left(\delta a-\delta
c\right)\right),\label{eq:StExpandedQuiverGH}
\end{split}
\end{equation}
and argue that the prescriptions (\ref{eq:cnaIndexExpanded}) are
independent of $G$,$H$,$C$, except that $H$ determines the pole
terms that according to the prescription of \cite{Ardehali:2015} one
should drop.

Following similar steps as in subsection \ref{app:B1} leads this
time to the coefficients
\begin{alignat}{3}
&a_0=G+H,\qquad && a_1=\frac{1}{2}G+\frac{1}{2}H+C, \qquad &&a_2=-\frac{8}{27}\delta a-\frac{8}{9}\delta c - \frac{1}{12}G-\frac{1}{12}H, \nonumber \\
&b_0=G,\qquad && b_1= \frac{3}{2}G,  \qquad
&&b_2=\frac{64}{27}\delta a- \frac{32}{9}\delta c+
\frac{1}{2}G.\label{eq:IstQuivCoeffs}
\end{alignat}
Importantly, this time $a_0-b_0=H\neq0$; this proves our claim that
the poles in (\ref{eq:cnaIndexExpanded}) (or alternatively, the
divergences encountered in \cite{Ardehali:2015,Ardehali:2014} for
large-$N$ theories) are due to the term proportional to
$\frac{1}{\beta(b+b^{-1})}$ in the high-temperature expansion of
$\ln \mathcal{I}(\beta,b)$.

Plugging the above set of coefficients in
(\ref{eq:cnaIndexExpanded}) leads to
\begin{eqnarray} \hat
a=\frac{9H}{32(t_t-1)^2}+a+\cdots,\quad \hat
c=-\frac{3H}{32(t_t-1)^2}+c+\cdots,
\end{eqnarray}
as expected, where again the dependence on $G$ has dropped out.
There is no dependence on $C$ either, as one could anticipate by
noting that $C$ enters only $a_1$ in (\ref{eq:IstQuivCoeffs}), while
$a_1$ does not appear in (\ref{eq:cnaIndexExpanded}).

Finally, we would like to point out that the above discussion
generalizes (up to the matching of the indices of bulk and boundary,
and assuming (\ref{eq:ESTfact})) the AdS/CFT matching of the
$\mathcal{O}(1)$ piece of the central charges to any toric quiver.
In \cite{Ardehali:2015} such a matching was demonstrated only for
toric quivers dual to smooth $SE_5$ manifolds and without adjoint
matter. The expression (\ref{eq:conjLargeNsq}) on the other hand
applies also to toric quivers with adjoint matter and with singular
dual geometry, as it hinges on the factorization (\ref{eq:ESTfact})
whose validity is demonstrated in several singular cases and in
presence of adjoints as well \cite{Agarwal:2013}.

The matching mentioned above could be alternatively demonstrated by
applying the prescriptions of \cite{Ardehali:2015} to the
single-trace index of a general toric quiver, which can be deduced
from Eq.~(\ref{eq:largeNquiverLogSq}) to be
\begin{equation}
I^{N\rightarrow\infty}_{s.t.\
quiver}(t,y)=-\frac{1}{2}\sum_{i=1}^{n_z} i_v(t_v=
t^{r_i},y_v=1)-\sum_{adj} i_\chi
(R=R_{adj},t_\chi=t,y_\chi=y),\label{eq:largeNquiverLogST}
\end{equation}
with $i_v(t_v,y_v)$ and $i_\chi(R,t_\chi,y_\chi)$ the single-letter
indices of the vector and chiral multiplets
\begin{equation}
i_v(t_v,y_v)=\frac{2t_v^2-t_v(y_v+y_v^{-1})}{(1-t_v y_v)(1-t_v
y_v^{-1})},\quad
i_\chi(R,t_\chi,y_\chi)=\frac{t_\chi^{R}-t_\chi^{2-R}}{(1-t_\chi
y_\chi)(1-t_\chi y_\chi^{-1})}.\label{eq:singleLett}
\end{equation}
Plugging the above explicit expressions in
(\ref{eq:largeNquiverLogST}) we obtain
\begin{equation}
I^{N\rightarrow\infty}_{s.t.\
quiver}(t,y)=\sum_{i=1}^{n_z}\frac{t^{r_i}}{1-t^{r_i}}-\sum_{adj}
\frac{t^{R_{adj}}-t^{2-R_{adj}}}{(1-ty)(1-t/y)},\label{eq:largeNquiverLogSimp}
\end{equation}
to which the prescriptions of \cite{Ardehali:2015} can be
successfully applied.

\end{document}